\documentclass{acm_proc_article-sp}
\usepackage{graphicx}

\begin{document}

\conferenceinfo{}{Bloomberg Data for Good Exchange 2016, NY, USA}

\title{An example of how false conclusions could be made with personalized health tracking and suggestions for avoiding similar situations}

\numberofauthors{2}
\author{
\alignauthor
Orianna DeMasi\\
       \affaddr{University of California, Berkeley}\\
       \affaddr{Berkeley, CA}\\
       \email{odemasi@berkeley.edu}
\alignauthor
Benjamin Recht\\
       \affaddr{University of California, Berkeley}\\
       \affaddr{Berkeley, CA}\\
       \email{brecht@berkeley.edu}
}

\maketitle
\begin{abstract}
Personalizing interventions and treatments is a necessity for optimal medical care. Recent advances in computing, such as personal electronic devices, have made it easier than ever to collect and utilize vast amounts of personal data on individuals. This data could support personalized medicine; however, there are pitfalls that must be avoided. We discuss an example, longitudinal medical tracking, in which traditional methods of evaluating machine learning algorithms fail and present the opportunity for false conclusions. We then pose three suggestions for avoiding such opportunities for misleading results in medical applications, where reliability is essential. 
\end{abstract}

%A category including the fourth, optional field follows...
%\category{D.2.8}{Software Engineering}{Metrics}[complexity measures, performance measures]

%\category{H.4}{Information Systems Applications}{Miscellaneous TODO}

%\terms{TODO}

%\keywords{TODO}

\section{Introduction}
The rapid development of computing and explosion in the amount of data collected has created myriad opportunities for medicine. These opportunities range from enabling entire fields, such as genomics and medical imaging to facilitating tasks, such as diagnosing conditions and monitoring chronic conditions. For many of these medical applications, machine learning algorithms are eliciting considerable interest \cite{hernandez14,hutson17, mukherjee17, parkin16}. There is great hope that algorithms and increased collection of data will increase the personalization of medicine. 

While medical opportunities are exciting in general, this excitement has the potential to lead to false optimism. Novel data sources and problem formulations that have emerged from medical applications bring with them the need for novel evaluation methods and assessment. While traditional methods of evaluation may be employed, they may be inappropriate for the application and result in false conclusions, which are at best expensive and at worst dangerous in medical applications. For example, researchers found that generic approaches for evaluating machine learning algorithms with cross-validation methods resulted in biased performance results \cite{saeb17} and such methods must be carefully applied on datasets with multiple individuals \cite{little17}. 

To mitigate poor or misleading evaluation of algorithms, interdisciplinary collaboration is needed to develop appropriate evaluation methods for medical applications. However, such collaboration may be difficult. Difficulty could arise from the common problems of insufficient communication and miscommunication between researchers who come from different communities. 

As an example of how standard evaluation methods may fail to hold algorithms accountable on applications that emerge from novel data sources, we consider the case of longitudinal health monitoring from smartphone data. This application has been popularly targeted as a candidate for machine learning algorithms \cite{piwek16}. Health monitoring is essential for understanding individuals' conditions and personalizing treatments \cite{chambless98, kazdin08}. While tracking acute conditions is easier, long-term tracking of chronic conditions has traditionally been a tedious practice requiring individuals to repeatedly manually record their personal state and thus has resulted in low compliance \cite{korotitsch99}. However, personal electronic devices have emerged as a novel data source of highly personal behavioral data that is easy to collect. As a result, many researchers hope such data may be a key to sustainably, automatically monitoring mental wellbeing without individuals constantly having to note their state \cite{benzeev15, canzian15, gruenerbl14, wang16}.

A standard evaluation method for an algorithm on any dataset might be to compare the percent of observations that the algorithm correctly predicted to the percent of observations that would have been correctly predicted just by assuming that all observations are the most frequently reported state. The insufficiency of this approach on the application of longitudinal health monitoring stems from the simple observation that individuals frequently report the same state, which can be different between individuals. Thus a better baseline for this application is to assume that each individual is always their personal most frequently reported state. We elaborate on this distinction between the former ``population baseline" with the latter ``personal baseline" below. When personal baselines are not used for comparison, false conclusions about the utility of an algorithm can be reached, but we find that personal baselines are not frequently used in a literature review. We use this example to highlight how misleading results can easily emerge from using non-specialized evaluation methods on a problem that has emerged from a new data source. 

With our experience working on health tracking in mind, we discuss efforts that may help improve results on this application. In particular, we suggest consideration of the following.
\\
First, studies could be made of failure modes or when health technologies should not be expected to help individuals. Such knowledge may increase the reliability of new technologies and highlight whether a technology is fairly serving subpopulations. Efforts to understand failure modes may be pursued with a data driven approach to identify whether a method may apply to an individual or with the development of new algorithms that reveal confidence (or lack thereof) in their predictions.

Second, a credit scheme could be developed to encourage and incentivize replicating results on sensitive but related datasets that cannot be publicly shared. Many personally identifiable datasets are collected by different institutions and cannot be shared to protect the participants. However, an incentive scheme where researchers could garner credit for reproducing results without gaining a novel research publication may encourage replication of results and comparison on different populations. 

Third, workshop sessions could be focused on developing and suggesting standards for evaluation methods and metrics for common problems. Novel datasets have enabled new problem formulations, that should critically be evaluated to avoid false conclusions. Interdisciplinary efforts are needed to evaluate methods from one domain applied to another and workshop sessions are an ideal place to do this. 

While we make these suggestions with health monitoring in mind, their application broadly relates to problems emerging from new datasets.

\section{Example of new data source}
Personal electronics, such as smartphones, smartwatches and fitness trackers are capable of collecting vast amounts of personal data on their users. Data are collected from the various sensors embedded in the devices that continuously stream data, e.g., accelerometers. Research has shown that these streams of sensor data relate to users' social relationships \cite{aharony11, eagle09} and that algorithms can be used to infer users' behaviors, such as sleep \cite{chen13} and activity \cite{incel13, lara13, lu10}. 

Some of the behaviors that smart devices can measure are thought to relate to mental wellbeing. As a result, researchers have begun to explore whether algorithms can also be used to infer mental wellbeing from behavioral data that are collected by smart devices \cite{mohr17}. In addition to diagnosing disorders, such as depression \cite{saeb15, saeb16}, there is hope that data from personal electronics can be used to monitor individuals' mental wellbeing, such as depression, stress, or states of other disorders, over time \cite{benzeev15, canzian15, gruenerbl14, wang16}. While interest in predicting and consequently monitoring state over time may not be new, the ready source of personal data from electronic devices has enabled this approach.

\subsection{Difference between population and \\personal baselines for evaluation}
%To explore whether smart devices can be used to monitor wellbeing, researchers often compare wellbeing predictions that are made from looking at data from the electronic device with  states that individuals reported during the course of the study. 
To evaluate whether an algorithm is any good at predicting wellbeing, its accuracy is quantified and then compared to the accuracy of a baseline, which essentially guesses answers rather than uses data to make predictions \cite{friedman01}. The baseline, should be tailored to the dataset. For example, the baseline for problems with rare events is to always guess that the rare event does not occur, as this is correct most of the time \cite{he09, weiss04}. For longitudinal monitoring, a baseline that guesses all individuals to always be at the same state, a ``population baseline", is misleading because individuals commonly report a single state, which differs between individuals. 

\begin{figure}[t] %  figure placement: here, top, bottom, or page
   \centering
   \includegraphics[width=\columnwidth]{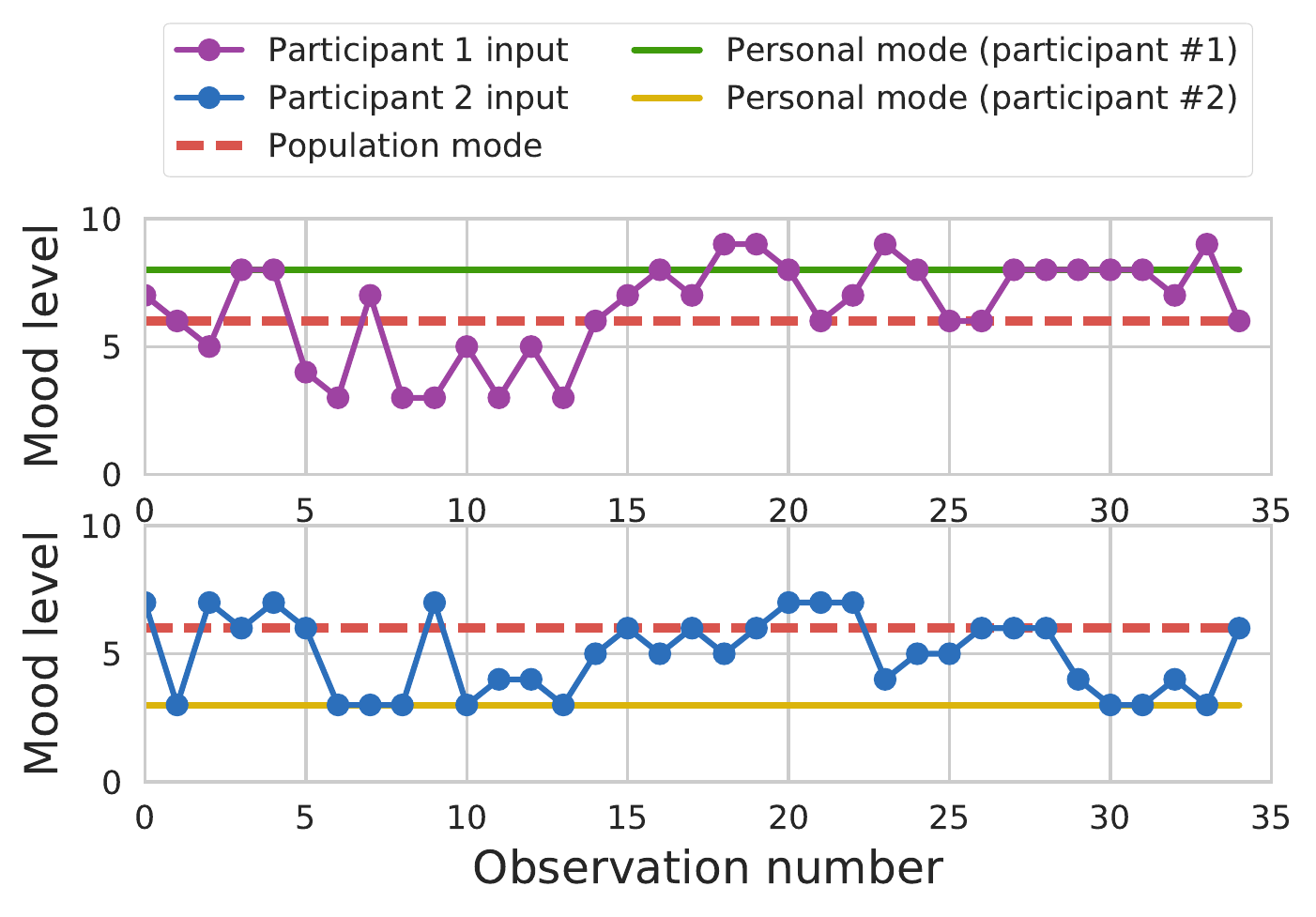} 
   \caption{Depiction of the difference between a personal mode, which is participant-specific, and a population mode, which might not represent the personal mode for any participant. Data are mood levels input on a smartphone app by two participants in a pilot study.}
   \label{fig:d4g_PopVsPerExample}
\end{figure}

An example of how individuals commonly report a single state that differs between individuals can be seen in Figure~\ref{fig:d4g_PopVsPerExample}. This figure shows consecutive mood observations that two individuals reported in a pilot study on nine-point Likert scales with higher values indicating a better mood. This data was collected by an app that sent a notification four times a day to each individual's smartphone that asked them what their mood and energy level were. Each individual had a commonly reported state, a ``personal baseline", but the states were different between participants.  In Figure~\ref{fig:d4g_PopVsPerExample} the population baseline, which was the most commonly reported state collected across both individuals' observations, was not representative of either individual. A similar observation holds when all the individuals in the study are considered -- the population baseline is not always representative of each individual's personal baseline.

\subsection{Hazard and prevalence of using incorrect baseline}

\begin{figure}[t] %  figure placement: here, top, bottom, or page
   \centering
   \includegraphics[width=\columnwidth]{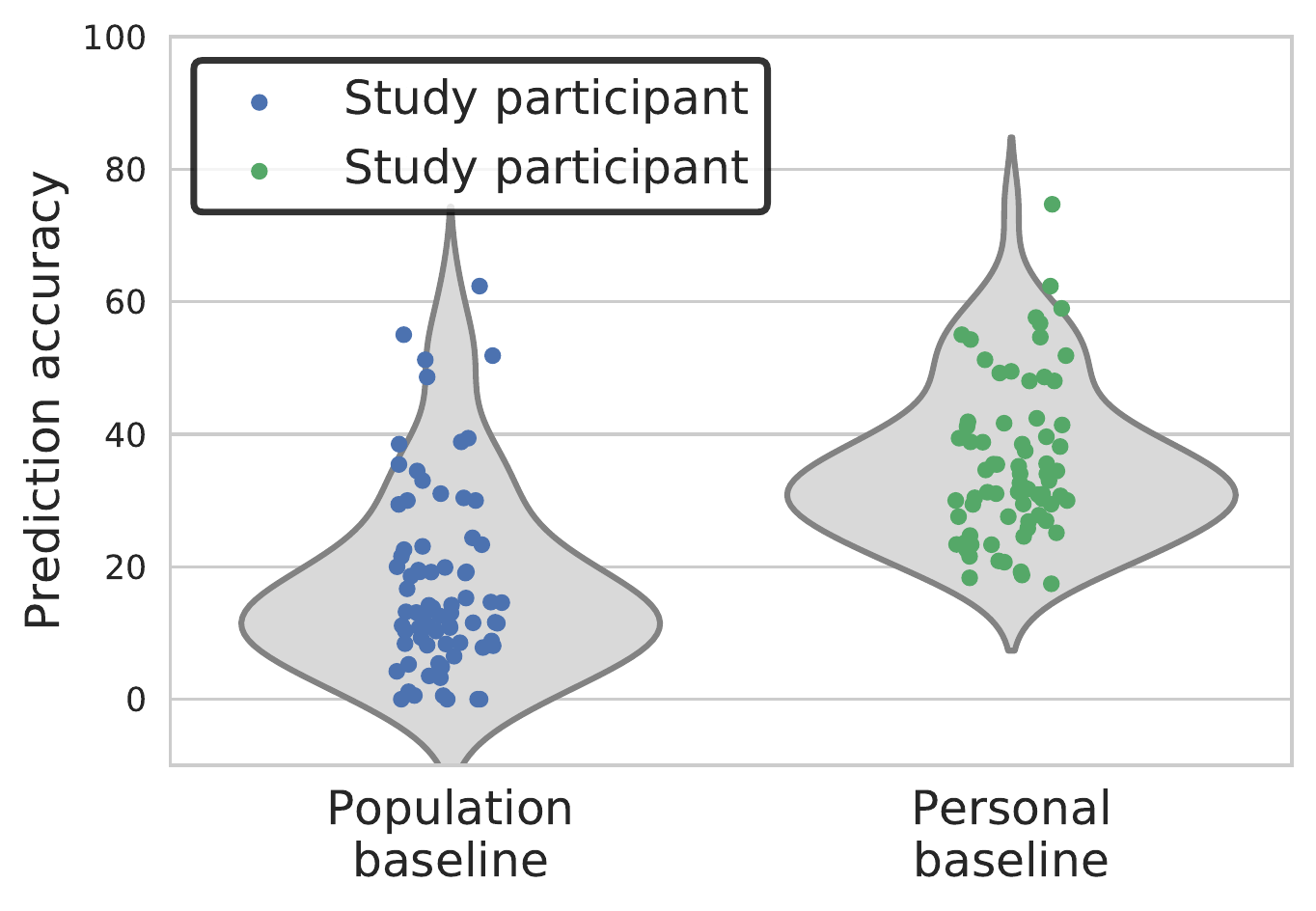} 
   \caption{Comparison of population and personal baseline accuracy for 73 participants' mood level inputs from a pilot study. Personal baselines can be considerably more accurate than population baselines. As a result, population baselines underestimate the accuracy a machine learning algorithm must have to be meaningful. Population and personal baseline accuracy are the percent of observations correctly predicted by always predicting each participant to always be the most commonly reported state within the population or within their own reports, respectively. }
   \label{fig:d4g_PopVPerBaseline}
\end{figure}

The distinction between a population and personal baseline is important, as an algorithm must be more accurate than both baselines, but showing that an algorithm is more accurate than a population baseline does not demonstrate that it is also better than a personal baseline. This result is because the population baseline accuracy, the percent of observations correctly predicted by guessing all individuals are always at the most commonly reported state in the population, can be lower than the personal baseline accuracy, the percent of observations that would be correct by guessing each individual to be at their personally most common state. The higher average prediction accuracy of personal baselines is demonstrated in Figure~\ref{fig:d4g_PopVPerBaseline}, by looking at personal and population baseline accuracy for mood input by 73 individuals from the above mentioned pilot study who had entered mood states. The personal baseline accuracy is higher on average across individuals than the population baseline accuracy, but both baselines are too simplistic to be meaningful.

Despite the need to show that algorithms are better than personal baselines, we found with a systematic literature review in previous work that a majority of studies do not compare with a personal baseline \cite{demasi17a}. This leaves the opportunity for studies to imply algorithms are good because they beat the (population) baseline, but in reality the algorithms may not be any better than always guessing each individual to be their most frequently reported state.

The lack of comparison with a personal baseline may be the result of the novelty of the application. There were previously limited sources of personal data that would make longitudinal data collection sustainable before smart devices. As a result, the application of algorithms for classification and regression to longitudinal wellbeing data is fairly new and thus the need for a new baseline, such as the personal baseline, that accounts for repeated measures is needed.

\section{Suggestions for progress on\\ health monitoring datasets}
The emergence of novel data sources and methods for processing said data, has created great opportunities for success, but also opportunities for false conclusions. The above sections outlined an example of how opportunities for false conclusions exist in one application, health monitoring. This opportunity resulted from comparing with a specific baseline, which was fine for traditional applications, but is insufficient for application to emergent longitudinal data sources. Similar opportunities for overly optimistic results have also been shown to emerge from insufficient care in other stages of evaluation, namely cross-validation when data on multiple individuals are considered \cite{saeb17}. However, there are actions, in addition to research directions, that may help avoid future opportunities for false positive conclusions and encourage progress within health monitoring.

\subsection{Identifying failure cases for increasing \\reliability and applicability}
As with many medical applications, it may be unrealistic to expect one solution for health monitoring to work for everyone. Various personalities, behaviors, and comfort levels with technology make mobile health (mHealth) technologies particularly likely to help some individuals more than others. Despite this expectation, there has been less emphasis in the above mentioned areas of smart device health monitoring, that we are aware of, on evaluating when such a technology may be applicable to an individual. 

\begin{figure}[t] %  figure placement: here, top, bottom, or page
   \centering
   \includegraphics[width=\columnwidth]{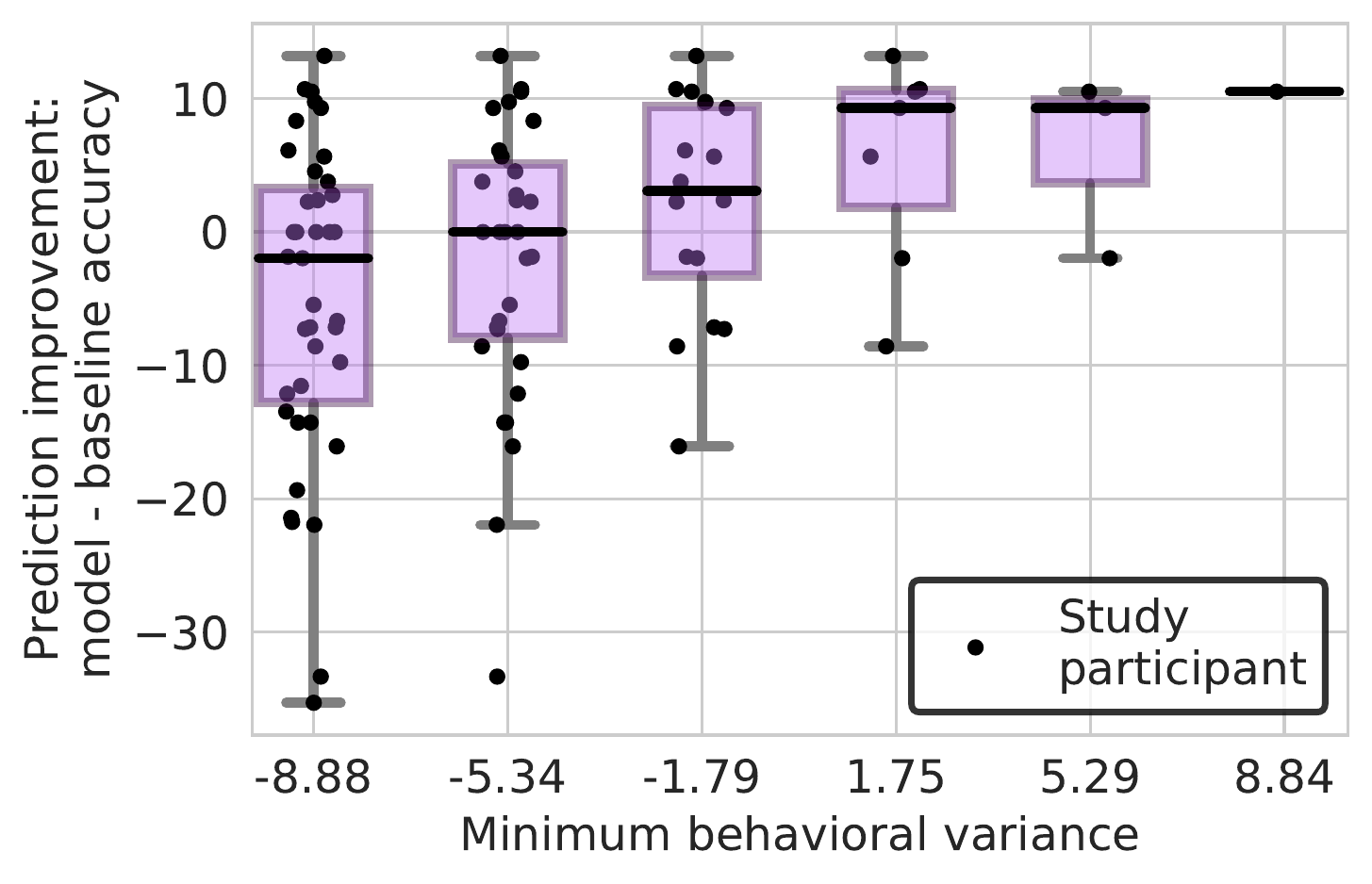} 
\caption{Average improvement in prediction accuracy when only individuals with a minimum level of behavioral variance are considered. Improvement in accuracy is higher for participants with higher behavioral variance, which may help identify participants who can and cannot be monitored by a smartphone. Models were constructed with location and mobility features from GPS logs and accuracy was from a random forest classifier predicting whether an individual's energy level was above average on a given day. }
   \label{fig:d4g_AccuracyVsLocVar_rf_rs}
\end{figure}

Rather than solely focusing on improving algorithms to always be accurate, reliability may be reached by identifying when the algorithms should not be expected to work. We have taken a  data driven approach and found possible evidence that improvement in prediction accuracy from using an algorithm is related to user behavior \cite{demasi17b}. In this work, we used classifiers to predict whether 39 participants (who had enough data to be included in the analyses) from the above mentioned pilot study were feeling more energetic than normal on a given day from previously described features quantifying location and mobility from their smartphone's GPS logs \cite{saeb15, saeb16}. We found that the improvement in prediction accuracy of simple classifiers over the personal baseline was positively correlated with behavioral variance, which was quantified by the log of the sum of variance of latitude and longitude from GPS logs during the eight-week study period. This result, while preliminary, indicates that individuals with less behavioral variance might be harder to predict. Figure~\ref{fig:d4g_AccuracyVsLocVar_rf_rs} shows how this finding might be used -- as participants are excluded from consideration, based on having too low behavioral variance, the average prediction improvement over a personal baseline increases. This might help identify for whom this technology will be accurate and for whom it will be too unreliable.

With future work, it might be possible to identify whether a monitoring approach will be feasible for individuals given features about them, such as their behavioral variance. Such analyses are important for increasing reliability, but also for identifying whether a technology may be unfairly underserving a population or missing the target population entirely. For example, if smartphones only sufficiently monitor mental wellbeing for active individuals, then alternative approaches may be needed for severely depressed individuals who may most need monitoring, but are often less active. 

In addition to studying when users might be predictable, developing algorithms that announce their certainty would be invaluable. These algorithms may take into account external data, such as a participant's age, or only their confidence based on previous data. 

\subsection{Database giving credit for reproducing \\results on sensitive datasets}

A pervasive research problem is the reproducibility of results, but this issue is even more challenging when collecting sensitive personally identifiable datasets. These sensitive datasets cannot be made publicly available for the protection of the study participants. However, as in the case of health monitoring, many institutions collect similar datasets and thus could potentially reproduce or compare results on similar populations. While there are examples of researchers attempting to reproduce results on related datasets \cite{asselbergs16, saeb17}, reproducing results on different datasets (and thus populations) is rarely done. This lack of reproduction is in part due to the lack of incentive scheme. 

Creating some centralized database and repository for registering studies, results, and sharing code, could generate such a scheme for credit assignment. With such a public system, junior researchers could garner credit for work they did to reproduce similar results on their own datatsets, without the need for the original researchers to collect another dataset and without the need for the junior researchers to find a publication venue for not fully unique results. 

While centralizing anything is difficult, projects have successfully helped researchers to coordinate. %For example, the Materials Project \cite{materialsproject} is an open database for known materials to accelerate discovery and understanding of materials. 
Further, many open-access journals have shown commitment to accessible and reproducible science and may be a natural shepherd for such an effort. With support, such venues may be able to have ``challenge areas" or sub-problems for which they create mini-repositories and databases where study details and results can be stored for comparison.

\subsection{Workshop sessions focusing on evaluation and problem formulation}

As illustrated with our example of longitudinal health monitoring, evaluation methods are problem specific. Interdisciplinary collaboration is needed to evaluate methods outside of their original domain and interdisciplinary workshop sessions may consider focusing some attention on evaluation methods and problem formulations. While the range of approaches and problem formulations that researchers take highlights a breadth of creativity that is necessary to solve challenging problems, it can also make comparing results between studies prohibitive. 

Leaders should hesitate to stifle the creative solutions that researchers use. However, having challenge problem formulations and evaluation suggestions that researchers can include to make their work comparable to other studies, could be invaluable additions and outcomes from the many workshops that have arisen \cite{ chiworkshop, clpsych, mhsiworkshop}. This approach has been taken by at least one workshop that has created a ``shared task" \cite{clpsych-sharedtask}, but broader suggestions of evaluation metrics could be envisioned.

\section{Conclusion}

We have discussed an application, health monitoring, where evaluation methods of of machine learning algorithms have been insufficiently applied, potentially because the application has recently been enlivened by novel data sources. This application highlights the need for interdisciplinary dialogue to define success. In light of our experience with health monitoring, we have highlighted three opportunities that may have facilitated developing a better method for evaluating monitoring solutions. These suggestions include identifying when and for whom methods fail, establishing a framework for assigning credit to researchers for reproducing results on similarly collected sensitive datasets, and focusing application specific workshop sessions on evaluation and challenge problem formulations for researchers to include as part of their studies. While these suggestions are targeted at a single application, health monitoring, they may be more broadly applicable to encouraging success with data science methods on novel datasets. 

%\nocite{*}
\bibliographystyle{abbrv}
\bibliography{D4G_bibliography}

\end{document}